\documentclass[12pt]{article}

\pdfoutput=1
\usepackage{amsmath,amssymb,exscale}
\usepackage{marvosym}
\usepackage{array,multicol}
\usepackage{afterpage,float,flafter}
\usepackage{amssymb}
\usepackage{epsfig,rotating,pifont}
\usepackage{cite}
\usepackage{hyperref}
\usepackage{color}
\usepackage{ marvosym }
\usepackage{textcomp}
\usepackage{wrapfig,slashed}

\definecolor{nicered}{rgb}{0.5,0.1,0.1}
\definecolor{nicegreen}{rgb}{0.1,0.5,0.1}
\definecolor{niceblue}{rgb}{0.1,0.1,0.8}
\hypersetup{colorlinks,citecolor= nicegreen,linkcolor= nicered}

\renewcommand{\Im}{\mathop{\rm Im}}

\newcommand{\mb}[1]{\mbox{\normalsize\boldmath $#1$}}

\@addtoreset{equation}{section} \makeatother

\def\beq{\begin{eqnarray}}
\def\eeq{\end{eqnarray}}
\def\lsim{\mathrel{\rlap{\lower3pt\hbox{\hskip0pt$\sim$}}
\raise1pt\hbox{$<$}}}         
\def\gsim{\mathrel{\rlap{\lower4pt\hbox{\hskip1pt$\sim$}}
\raise1pt\hbox{$>$}}}         

\def\SU{{\rm SU}}
\def\SO{{\rm SO}}

\def\circa#1{\,\raise.3ex\hbox{$#1$\kern-.75em\lower1ex\hbox{$\sim$}}\,}

\newcommand{\GeV}{\,{\rm GeV}}
\newcommand{\TeV}{\,{\rm TeV}}

\font\tenrsfs=rsfs10 at 12pt

\font\sevenrsfs=rsfs7
\font\fiversfs=rsfs5
\newfam\rsfsfam
\textfont\rsfsfam=\tenrsfs
\scriptfont\rsfsfam=\sevenrsfs
\scriptscriptfont\rsfsfam=\fiversfs
\def\mathscr#1{{\fam\rsfsfam\relax#1}}
\def\Lag{\mathscr{L}}

\title{
\vspace{-3cm}
\begin{center}
\small{CERN-PH-TH/2012-233}
\end{center}
\vspace{1.7cm}
\begin{center}
\medskip
{\Huge\bf 
Axion-Higgs Unification
}
\end{center}
\vspace{0.6cm}
\author{
\Large{\text{\bf Michele Redi$^{1,2}$}\footnote{michele.redi@cern.ch}~~and \text{\bf Alessandro Strumia$^{3,4}$}\footnote{alessandro.strumia@cern.ch}~~
}
\\ \\
$^1$\emph{CERN, Theory Division, CH-1211, Geneva 23, Switzerland}\\
$^2$\emph{INFN, 50019 Sesto Fiorentino, Firenze, Italy}\\
$^3$\emph{Dipartimento di Fisica dell'Universit{\`a} di Pisa and INFN, Italy}\\
$^4$\emph{National Institute of Chemical Physics and Biophysics, Tallinn, Estonia}\\
}
}

\date{}
\begin{document}
\maketitle \thispagestyle{empty} \vspace*{-.2cm}

\centerline{\large\bf Abstract}
\begin{quote}\large
In theories with no fundamental scalars, 
one gauge group can become strong at a large scale $\Lambda$ and
spontaneously break a global symmetry, 
producing the Higgs and the axion as composite pseudo-Nambu-Goldstone bosons.
We show how KSVZ and DFSZ axion models can be naturally realised.
The assumption $\Lambda\approx 10^{11} \GeV$ is phenomenologically favoured  because:
a) The axion solves the QCD $\theta$ problem and provides the observed DM abundance;
b) The observed Higgs mass is generated via RGE effects
from a small Higgs quartic coupling at the compositeness scale,
provided that the Higgs mass term is fine-tuned to be of electroweak size;
c) Lepton, quark as well as neutrino masses can be obtained from four-fermion operators at the compositeness scale.
d) The extra fermions can unify the gauge couplings.
\end{quote}

\newpage
\renewcommand{\thepage}{\arabic{page}}
\setcounter{page}{1}

\section{Introduction}
Present flavour data, precision data, LHC data are compatible with the Standard Model (SM)  and no new physics.
We will assume that this means that the naturalness criterion adopted as guideline by many
theorists in  the past 30 years is  wrong.  

Furthermore, the measured  Higgs mass $m_h = 125.5\pm 0.5$ GeV~\cite{exp}
corresponds to a Higgs quartic coupling $\lambda$ close to zero when renormalized
at energies above $\Lambda \approx 10^{11}$ GeV~\cite{RGESM}.
A small quartic $\lambda$ can have alternative theoretical interpretations
in terms of new physics at the scale $\Lambda$, such as
\begin{itemize}
\item[1)] Supersymmetry with $\tan\beta=1$ broken at $m_{\rm SUSY} \approx \Lambda$~\cite{Giudice,hebecker,ibanez};
\item[2)] The Higgs could be the 5-th component of a gauge field living
in one extra dimension with radius $R\approx \Lambda^{-1}$;
\item[3)] The Higgs could be the Nambu-Goldstone-Boson (NGB) of a global  symmetry $G$ spontaneously broken 
to a subgroup $H$ at some high scale $f\approx \Lambda$. 
\end{itemize}
We here explore the third possibility that finds a natural realization in theories
where one extra `Techni-Color' (TC) gauge interaction becomes strong at the scale $\Lambda$~\cite{TCreviews}.\footnote{Other realizations 
of the same idea are in weakly coupled theories, or in models with extra dimensions (warped or not warped) 
where the Higgs is the Wilson line of an extra-dimensional  gauge field \cite{minimalcomposite}.}
We are interested in Composite Higgs (CH) models~\cite{georgi}, where condensates break a global symmetry producing the Higgs as 
a pseudo-NGB,  while the SM gauge group is not broken by the strong dynamics.
One general feature is  a small Higgs quartic coupling at the TC scale.  The quartic coupling responsible for the
observed Higgs mass is then generated  dominantly via SM RGE running down to the weak scale.
\medskip

Composite Higgs models, already discussed at the weak scale $\Lambda \sim \TeV$ as possible solutions to the hierarchy problem, 
have various problems: no new physics has so far shown up in experiments
such that one needs to resort to ad-hoc constructions with no known UV completion.
All these problems disappear if $\Lambda \sim 10^{11}\GeV$, and furthermore
RGE corrections enhanced by $\ln \Lambda/v$ generate the Higgs potential
even starting from $V(h)=0$ at tree level.
One has to accept a huge fine-tuning of the electro-weak scale $(\Lambda/v)^2\sim 10^{18}$, 
possibly explained by `anthropic selection'~\cite{Don}.

%

\medskip

Another motivation for this work is that TC models tend to automatically predict axion candidates~\cite{TCreviews}.
If $\Lambda\sim\TeV$  this is an embarrassment ---   weak scale axions are excluded by
experiments, while Dark Matter considerations favour a much higher axion scale, $f_a\approx 10^{11}$ GeV~\cite{axionDM} ---
so that appropriate mechanisms must be invoked to remove weak-scale axions.
The natural occurrence of axions becomes a virtue in our scenario where the TC scale is around $10^{11}\GeV$: 
such models can simultaneously explain the QCD $\theta$ problem and Dark Matter and  lead to  predictions for the axion/photon couplings.

Furthermore, if these models could be extended into models for fermion masses following the old idea
of Extended Technicolor (ETC)~\cite{TCreviews}, neutrino masses can naturally arise with the observed order 
of magnitude, $m_\nu\sim v^2/\Lambda\sim {\rm eV}$.

\medskip

In this paper we do not attempt such ambitious ETC construction, and explore
models where the axion and Higgs are unified in an irreducible coset $G/H$.
They can be  divided in two categories, depending on how the QCD anomaly of the axion symmetry is generated:
\begin{itemize}

\item[A)]  from {\em new fermions} (KSVZ axions).
Within models where the Higgs and axion are unified in a single coset $G/H$ this possibility
arises when the full SM gauge group is contained in the unbroken group $H$. This can realize
KSVZ axions \cite{KSVZ}.

\item[B)] from {\em the SM fermions} (DFSZ axions).
This possibility arises when color $\SU(3)_c$ is not contained in the unbroken group $H$.
Since there is no contribution to the QCD anomaly from TC fermions the anomaly must be generated 
by the SM fermions. The composite sector can however contribute to the electro-magnetic anomaly, 
because the unbroken group $H$ contains the electro-weak group. This is similar to DFSZ axions \cite{ZDFS}.

\end{itemize}

Another distinction arises because, within each class, the SM fermion masses can either be generated: 
\begin{itemize}
\item[a)]  from {\em bi-linear} couplings of SM fermions to TC fermions. This possibility can be presumably realized 
in appropriate ETC constructions that unify the SM gauge group and TC interactions.

\item[b)] from {\em linear} couplings of SM fermions to fermions of the TC sector.
This possibility, known as partial compositeness,
is often considered in recent attempts  of building weak-scale composite Higgs models compatible with data, 
but it seems harder to realize this in known UV completions.

\end{itemize}

KSVZ models  are explored in section~\ref{Aa};
DSFZ models in section~\ref{Ba}; models with partial compositeness in section \ref{linear}. 
The Higgs mass in  models with high compositeness scale is discussed in section \ref{higgsmass}. 
In section~\ref{concl} we conclude. Technical details and estimates of the Higgs potential are 
collected in the appendix.

\section{Models with composite Higgs and KSVZ axion}
\label{Aa}

First, let us recall that KSVZ axion models~\cite{KSVZ} add to the SM vector-like heavy quarks $\Psi_Q=(Q_L, \bar Q_R)$
\footnote{Here and in what follows all fermions are two-components left-handed Weyl fermions.} and a
heavy complex scalar $\sigma$ with a Peccei-Quinn (PQ) symmetry 
\beq \Psi_Q \to e^{i \alpha_Q}\Psi_Q,\qquad \sigma\to e^{-2i\alpha_Q}\sigma  \eeq
that forbids the $\Psi_Q$ mass term.
Heavy quark masses only come from the  VEV of $\sigma$ that spontaneously breaks the chiral symmetry:
\beq \Lag = \Lag_{\rm SM} + \bar\Psi_Q i\slashed{\partial} \Psi_Q + |\partial_\mu \sigma|^2+
(\lambda\,\sigma\, \bar\Psi_Q \Psi_Q  +\hbox{h.c.}) - V(\sigma)\eeq
giving rise to a light axion $a=\sqrt{2} \Im\sigma$ with large decay constant $f_a \approx \langle\sigma\rangle$.

\medskip

Composite Higgs models that realize the KSVZ axion can be broadly characterised as follows.
The broken generators of $G/H$ contain the PQ symmetry; the unbroken group
$H$ contains the SM gauge group and the NGB include the Higgs doublet and the axion singlet.
To solve the strong CP problem the axionic symmetry should be anomalous under 
QCD but not under TC gauge interactions.

Given that the SM gauge group is a subset of SU(5), we find that a minimal possibility  is provided by\footnote{$\SU(5)_L\times \SU(5)_R /\SU(5)_{L+R}$ would not work because the NGBs do not include doublets.}
\begin{equation}
\frac{G}{H}=
\frac {\SU(6)_L\times \SU(6)_R}{\SU(6)_{L+R}}\ .
\label{chiralsu6}
\end{equation}
The 35 NGBs decompose under $\SU(5)_{\rm SM}$ as ${\bf 24}\oplus {\bf 5}\oplus \bar {\bf 5} \oplus {\bf 1}$.
The 33 scalars charged under the SM interactions typically acquire a large mass of order $g_{\rm SM} \Lambda/4\pi$.
We assume that the scalar with the quantum numbers of the Higgs boson remains accidentally light presumably due 
to `anthropic cancellations' in the potential, see section~\ref{higgsmass}.
The two scalar SM singlets are axion candidates.

\subsection{A simple model}
To be concrete, the above scenario can be realized adding to the SM an $\SU(n)$ TC group
and massless TC-fermions transforming in the following representations of the full gauge group,
${\rm U}(1)_Y\otimes\SU(2)_L \otimes \SU(3)_c \otimes \SU(n)_{\rm TC}$:
\beq
\begin{array}{ccccc}\hline
\hbox{Fermions}&{\rm U}(1)_Y&\SU(2)_L&\SU(3)_{\rm c} & \SU(n)_{\rm TC}\cr \hline
D  & \phantom{-}{1 \over 3}& 1 &\bar{3}  & n\cr
L & -{1 \over 2} & 2 &1 & n \cr
N & \phantom{-}0 & 1 & 1 & n\cr
\bar D  &-{1 \over 3}& 1 &3  & \bar n\cr
\bar L & \phantom{-}{1 \over 2} & \bar 2 &1 & \bar n \cr
\bar N & \phantom{-}0 & 1 & 1 & \bar n
\end{array}
\eeq
Massless TC fermions can be enforced with a Z$_2$ symmetry.
Notice that we do not make use of any fundamental scalar:
$D, L$ and $N$ are TC-colored copies of the SM fermions (right-handed down quarks $d$,
left-handed lepton doublet $\ell$, right-handed neutrino $n$), and they fill full SU(5) multiplets $\bar 5 \oplus 1$.
The main difference with respect to TC constructions is the addition of fermions
 $\bar D,\bar L$ and $\bar N$ in the conjugated representations, such that TC fermions 
are vectorial with respect to the SM gauge group. 
This guarantees that the TC dynamics does not break the electro-weak symmetry.
Indeed the following condensates develop
\begin{equation}
\langle D \bar{D} \rangle = \langle L \bar L\rangle =\langle N\bar{N} \rangle \approx  \Lambda^3\, 
\end{equation}
realizing the pattern~(\ref{chiralsu6}) of symmetry breaking.
Among the 36 scalars $D\bar D, L\bar L, N\bar N$, $D\bar L$, $L\bar D, D\bar N, N\bar D, L\bar N, N\bar L$
there is a single NGB Higgs doublet that corresponds to the combination
\begin{equation}
H\sim (L \bar{N})- ({\bar L} N)^*
\label{higgses1}
\end{equation}
and three neutral real states: $D\bar D, L\bar L$ and $ N\bar N$.
One of them, $D\bar D+L\bar L+ N\bar N$ is not a true NGB: it gets
a mass of order $g_{TC} \Lambda/4\pi$ from the ${\rm U}(1)\times \SU(n)^2_{\rm TC}$ anomaly, just like the $\eta'$ in QCD.
The orthogonal combinations typically have a QCD anomaly and 
provide viable axion candidates.

\medskip

Their precise phenomenology depends on model options not specified so far. Let us summarize the main possibilities.

\begin{enumerate}
\item
If high-scale dynamics respects their global symmetries,
at low energy one has two light NGB associated to the U(1) charges,
\beq
 \frac {4D - 3L -6 N} {\sqrt{102}},\qquad
\frac {L-2 N} {\sqrt{3}} \ .
\eeq
The first one has a QCD anomaly and behaves as an axion with electromagnetic coupling 
$E/N = -5/6$.\footnote{
We recall that it is defined as
$E/N=\sum Q_{\rm PQ} Q_{\rm em}^2/\sum Q_{\rm PQ} T^2$ where the sums extend over all fermions with PQ charges $Q_{\rm PQ}$, electric charges $Q_{\rm em}$, and QCD generators
${\rm Tr}\, T^a T^b = \frac{1}{2}T^2\delta^{ab}$.
The  electromagnetic coupling of the axion is predicted as
$$
\frac{g_{a\gamma\gamma}}{m_a} = \frac{\alpha_{\rm em}}{2\pi f_\pi m_\pi}\sqrt{(1+\frac{m_d}{m_u})(1+\frac{m_u}{m_d}+\frac{m_u}{m_s})}
\left[ \frac{E}{N}-\frac{2}{3}\left( \frac{4+m_u/m_d+m_u/m_s}{1+m_u/m_d+m_u/m_s}\right) \right] .$$
where $f_\pi = 93$ MeV.
Thereby $E/N$ can be measured up to a two-fold ambiguity.
Axion dark matter experiments such as ADMX~\cite{ADMX} are reaching now the required sensitivity
for the value of $f\approx 10^{11}\GeV$ suggested by the cosmological DM abundance.
}
The presence of the other NGB does not pose a phenomenological problem, 
because its interactions are suppressed by $f\sim10^{11}\GeV$~\cite{Savas}. 

\item
If instead the SM gauge group
is unified into SU(5) at a scale $M_{\rm GUT}$, then $D$ and $L$ merge into a $\bar 5$ such
that their phases cannot be independently rotated any more, and one of the two NGB gets 
a mass of order $\Lambda^2/M_{\rm GUT}$. The NGB that survives at low energy and plays the role of the axion
is the SU(5)-invariant
\beq \frac{D + L -5 N}{\sqrt{30}} \eeq
with $E/N=8/3$.\footnote{
The fact that the Higgs doublet is charged under the symmetry associated to the axion is not a problem
and does not affect their phenomenology, because one can compensate a change in the Higgs phase by a 
U(1)$_{Y - B/6}$ rotation of the SM fields. This transformation introduces no QCD nor QED anomalies.}

\item
SM Yukawa couplings presumably originate from 4-fermion operators, which
could break none, some or all of the axion-like symmetries.
(The latter option is the one pursued in weak scale TC models).
For example, the Yukawa couplings for the SM up-type quarks $q,u$ can be generated by two kind of operators:
\beq  (q u) (L\bar N),\qquad (\bar q \bar u)(\bar L N)
\label{yuk} \eeq
If both of them are present, then one NGB gets a large mass of order $\approx y_t \Lambda$
and the surviving light axion is the combination under which the Higgs doublet is neutral:
\beq\frac {D -3 L + 3 N} {\sqrt{30}}
\eeq
In such a case, the model predicts the experimentally observable 
ratio of axion electro-magnetic and QCD anomalies to be $E/N=-16/3$.
\end{enumerate}

Models where the only light composite particle is one axion (but not the Higgs) can be easily built 
along the same lines, overcoming the difficulties in~\cite{Sannino}.

\smallskip

Of course, one can build variants of the model with different TC-fermion content.
For example the TC fermions might fill a full family multiplet $(\bar {\bf 5} \oplus {\bf 10}  \oplus {\bf 1}, \mb{n})+\hbox{h.c}$.
The global symmetry is now $G/H=\SU(16)_L\times \SU(16)_R/\SU(16)_{L+R}$. 
This is the composite Higgs analog of the one family TC model of Farhi-Susskind \cite{farhisusskind}. 
This model contains 7 axion candidates. The one coupled to $5 \cdot {\bf 5}-2 \cdot {\bf 10}-5 \cdot {\bf 1}$ 
predicts $E/N=4/3$ and is particularly interesting, because it combines the good properties
previously discussed: the Higgs has no PQ charge, and it is SU(5)-invariant.

\smallskip

Alternatively, TC fermions might fill a `tilted' family like $({\bf 5} \oplus {\bf 10} \oplus {\bf 1}, \mb{n})+\hbox{h.c.}$: 
such models lead to extra states with the quantum numbers of the Higgs doublet  (like $LE$, $QD$).

\begin{figure}[t!]
\begin{center}
$$\includegraphics[width=0.45\textwidth]{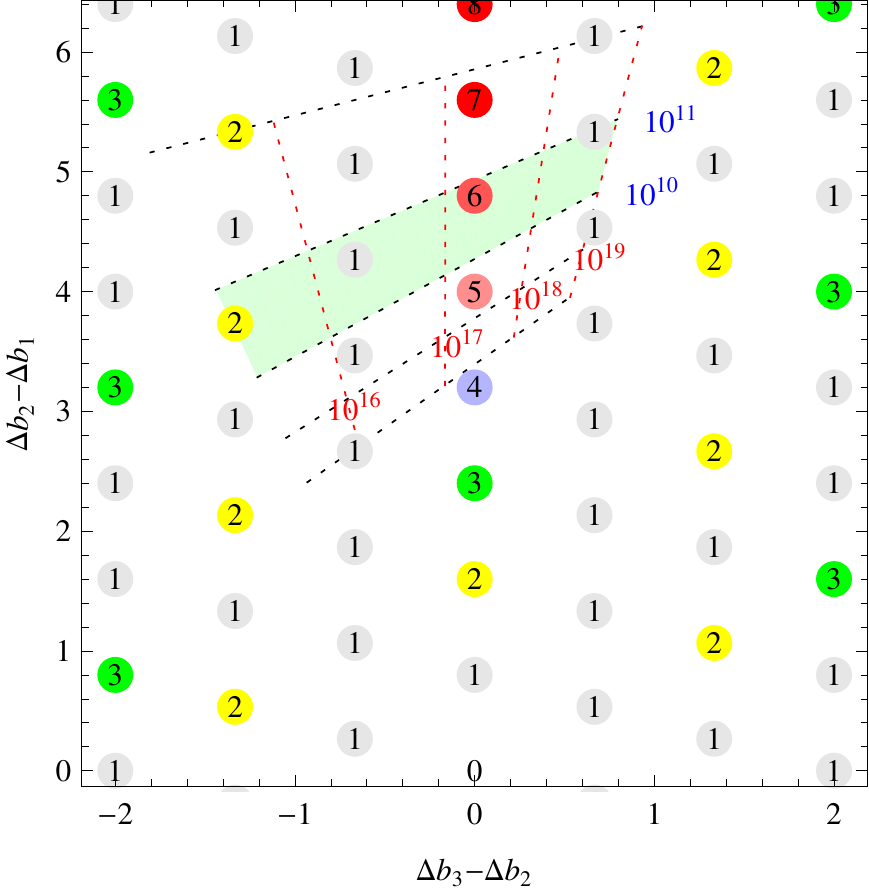}\qquad
\includegraphics[width=0.45\textwidth]{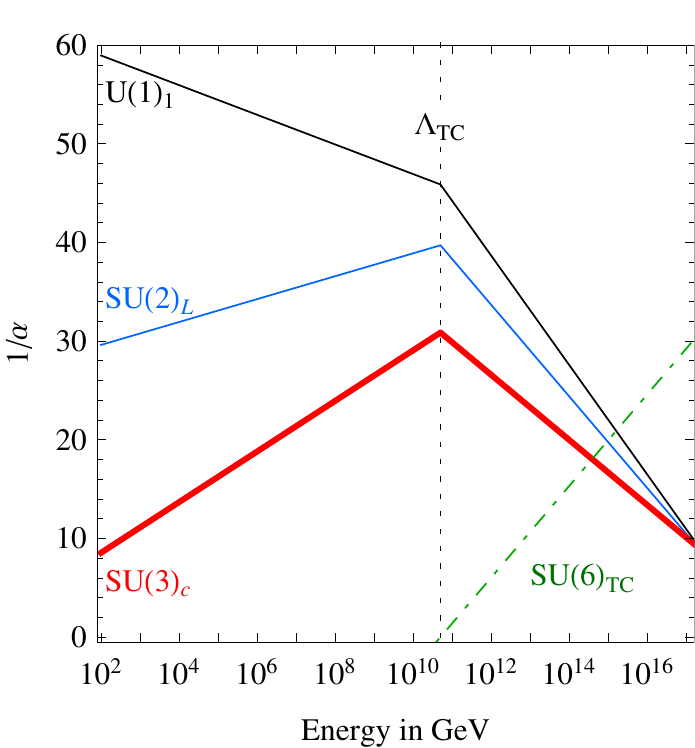}$$
\caption{\em {\bf Left:}
Possible values of the one-loop $\beta$-function coefficients generated by fermions in real representations of the SM group that can be embedded in {\rm SU(5)} multiplets marked according to their  maximal multiplicity
(e.g.\ the point marked with $6$ can be realised with up to 6 copies of the same set of fermions).
The shading indicates the region favoured by unification
with multiplets of  mass $\Lambda$ between
$10^{10}$ and $10^{11}\GeV$ (the GUT scale in GeV is indicated in red). {\bf Right:} Running of the gauge couplings 
(with hypercharge in GUT normalization)
for one model that 
 realises the point marked with $6$ employing a $\SU(6)_{\rm TC}$ TC interaction.
}
\label{nGUTs}
\end{center}
\end{figure}

\bigskip

In the model above the presence of strong TC dynamics at $\Lambda \approx 10^{11}\GeV$
does not improve the unification of the SM gauge couplings, as the composite states come in complete multiplets. 
This remains true at any order (even non-perturbative) in the TC coupling.

\smallskip

So, one can instead consider  TC fermions in incomplete SU(5)$_{\rm GUT}$ multiplets.
Along the lines of~\cite{unificaxion},
fig.~\ref{nGUTs}a shows the values of the 
one-loop $\beta$-function coefficients for the SM gauge couplings
allowed by arbitrary
combinations of fermions in real representations of the SM group that can be embedded in {\rm SU(5)}$_{\rm GUT}$ multiplets.
Within the shaded region, unification is achieved
with an intermediate mass $\Lambda$
between $10^{11}$ and $10^{12}\GeV$.\footnote{Such region
slightly differs from the analogous figure in~\cite{unificaxion} because in the present scenario
the Higgs boson is not present above the compositeness scale.} 
We added on each point 
a number $n$ equal to the maximal multiplicity of the fermions.
This means that such point can be realised adding $n$ copies of the same set of fermions.
This issue is relevant in the TC context, where TC fermions appear in the $\mb{n}\oplus\bar{\mb{n}}$ representation
of a $\SU(n)$ TC group.
This singles out the points marked as $n=6$ and $n=5$ around the shaded region favoured by unification.
Such points (where unification occurs at $M_{\rm GUT} \approx 2\times 10^{17}\GeV$)
can be  obtained adding TC fermions in 
an incomplete SM family $D,L,Q,U,N$ (no right-handed leptons $E$) 
in the
$\mb{n}$ representation of $\SU(n)_{\rm TC}$,
plus their conjugated $\bar D,\bar L,\bar Q,\bar U,\bar N$ in the $\bar{\mb{n}}$.
The corresponding evolution of the gauge couplings is shown in fig.~\ref{nGUTs}b for the case $n=6$.

\section{Models with composite Higgs and DFSZ axion}
\label{Ba}
First, let us recall {Dine-Fischler-Srednicki-Zhitnitskii  (DFSZ) axion models}~\cite{ZDFS}. They use,
instead of extra colored fermions, two Higgs doublets and one complex heavy neutral scalar $\sigma$. 
The theory is assumed to be symmetric under a global U(1)$_{\rm PQ}$  
\beq  \sigma\to e^{4i\alpha}\sigma,\qquad
q_{L,R}\to e^{i\alpha }q_{L,R}\qquad H _u \to e^{-2i\alpha} H_u,\qquad H_d \to e^{-2i\alpha} H_d \eeq
such that a term $\sigma H_u H_d$ is allowed and present in the Lagrangian,
while $H_u H_d$ and $(H_u H_d)^2$ terms are absent. 
The PQ symmetry is then broken by the vacuum expectation values of $H_u,H_d$ and $\sigma$.
Assuming that $\sigma$ has a large vev $\langle\sigma\rangle\gg v$ the axion is $a = \sqrt{2}\Im\sigma$ with large
decay constant $f_a \approx \langle \sigma\rangle$.

\medskip

Composite Higgs models that realize this scenario can be characterised as follows.
The group $G$ contains the PQ symmetry;
the unbroken group $H$ contains the EW group (but not the QCD group),
the NGB contain the axion and two Higgses, that must be charged under the PQ symmetry.

\subsection{A simple model}

In this section we describe a realization of this scenario based on the pattern 
\beq \frac{G}{H}=\frac{\SU(6)}{\SO(6)}\label{eq:SO(6)}
\eeq
that generalizes the  model of Kaplan and Georgi \cite{georgi}
such that $\SO(6)\supset \SO(4)\otimes {\rm U}(1)_{\rm PQ}$,
where $\SO(4)  \supset \SU(2)_L\otimes\SU(2)_R$ contains 
the SM electroweak group (hypercharge is the $T_3$ generator of $\SU(2)_R$).

The pattern of symmetry breaking in eq.~(\ref{eq:SO(6)}) can be obtained from a $\SO(n)_{\rm TC}$ gauge theory 
with 6 massless TC-fermions $\psi^i$ in the vectorial representation of the TC gauge group \cite{witten}. They can be organized as
\beq \begin{array}{cccccc}\hline
\hbox{Fermions}&{\rm U}(1)_Y&\SU(2)_L&\SU(3)_{\rm c} & \SO(n)_{\rm TC}& {\rm U}(1)_{\rm PQ}\cr \hline
L & -{1 \over 2} & 2 &1 & n&\phantom{-}0 \cr
\bar L & \phantom{-}{1 \over 2} & \bar 2 &1 &  n &\phantom{-}0\cr
N & \phantom{-}0 & 1 & 1 & n & \phantom{-}2\cr
\bar N & \phantom{-}0 & 1 & 1 &  n&-2
\end{array}\eeq
where $L=(\psi_3+i \psi_4 ,\psi_1-i \psi_2)$, $\bar{L} =(\psi_1+i \psi_2 , -\psi_3+i \psi_4)$, $N=\psi_5+i \psi_6$ and $\bar{N}=\psi_5-i \psi_6$ are the combinations with definite PQ charge.

The TC condensate $\langle L\bar L\rangle = \langle N\bar N\rangle= \Lambda^3$ breaks $\SU(6)\to\SO(6)$.
The NGB, loosely written as 
$$  LN, 
L \bar N, \bar L N,\bar L \bar N ~\oplus~
  NN, \bar N \bar N~\oplus~ N\bar N~\oplus~ LL ,L\bar L, \bar L\bar L,$$
transform as the ${\bf 20}'$ (symmetric traceless) representation
of the unbroken $\SO(6)$, that decomposes under $\SU(2)_L\otimes\SU(2)_R\otimes {\rm U}(1)_{\rm PQ}$ as
\begin{equation}
{\bf 20'}=  \bf{(2,2)_{\pm 2}} \oplus \bf{(1,1)_{\pm 4}} \oplus \bf{(1,1)}_0  \oplus \bf{(3,3)}_0 
\label{20prime}
\end{equation}
where the subscripts show the PQ charge. The first two multiplets charged under U(1)$_{\rm PQ}$
are the scalars $H_u,H_d$ and $\sigma$ of the DFSZ axion models;
furthermore there is a singlet $\eta$ and  scalars with EW gauge interactions
that will acquire a large mass. We again assume that one Higgs doublet remains anthropically light.

\medskip

The potential $V(\sigma,H_u,H_d)$ (which needs to have a minimum at $\sigma \approx \Lambda$) is generated only by interactions 
that break the global symmetry: in the minimal case these are gauge interactions and four-fermion operators needed to produce 
the SM Yukawas.
We assume that such interactions respect the PQ symmetry, realizing a type-II
two Higgs doublet model:
\begin{equation}
\frac {1}{\Lambda^2_t}  (q_L t_R^c)^\dagger (L\,   N) + \frac {1}{\Lambda^2_b}  (q_L b_R^c)^\dagger (\bar{L}\,   \bar{N}) + \hbox{h.c.}
\label{dim6DFSZ}
\end{equation}
When converted into couplings to the NGBs $H_u$ and $H_d$, couplings to $H_u$
are accompanied by couplings to $\sigma^{*} H_d^*/f^2$, such that these interactions generate at
one loop (see appendix~\ref{A2app}) the $\sigma H_u H_d$ term needed for the DFSZ axion mechanism.

The ratio $E/N$ is uniquely determined in these models. 
There are no anomalies associated to the composite sector fermions so one finds the same results of \cite{ZDFS},
\begin{equation}
\frac {E}{N}=\frac 8 3.
\end{equation}

\medskip

Let us briefly discuss neutrino masses. If the SM fields include the right-handed neutrino we can generate a small
mass with the standard see-saw mechanism. Alternatively, compatibly with PQ symmetry, we can add the dimension 9 operator,
\begin{equation}
\frac {1} {\Lambda_\nu^5}\, (\ell \bar L)^2 N^2\to
\frac {1} {\Lambda_\nu^3}(\ell H_u)^2 \sigma^2 + \cdots
\end{equation}
that gives rise to Majorana masses of order $v^2/ \Lambda$ after that $\sigma$ acquires 
a VEV of order $f$.

\section{Partially Composite Axion-Higgs}\label{linear}
In this section we discuss  the most minimal model of Axion-Higgs unification, where the NGB are just the $4+1$ degrees of freedom
of the Higgs doublet and of the  axionic phase.
Such scenario is realized by the coset
\beq \frac{G}{H}=\frac{\SO(6)}{\SO(5)}\simeq \frac{\SU(4)}{{\rm Sp}(4)}
\eeq
where $\SO(6)$ contains the EW gauge group and the PQ symmetry as
$\SO(4)\otimes\SO(2)= \SU(2)_L\otimes \SU(2)_R\otimes {\rm U}(1)_{\rm PQ}$~\cite{singlet}. 
The 5 NGBs, transforming in the {\bf 5} of $\SO(5)$, decompose into a doublet and a singlet under the electro-weak symmetry. 
The NGB can be parametrized with an $\SO(6)$ vector with unit norm that 
in the unitary gauge can be written in terms of the physical Higgs
$h$ and axion $a$ as:
\begin{equation}
\Phi=\left(0,0,0,\frac{h}{f} ,
\frac{a}{f} ,\
\sqrt{1-\frac{h^2+a^2}{f^2}}\right).
\end{equation}
The U(1)$_{\rm PQ}$, which acts as rotation between the last two components of $\Phi$
is spontaneously broken by 
the strong dynamics at $\Lambda$ so that the EW singlet $a$ is the axion candidate.\footnote{
This was already noted in \cite{singlet} where the same coset was considered. There however $f\sim v$ so that ${\rm U}(1)_{\rm PQ}$ had to be 
explicitly broken to remove the unwanted weak-scale axion. We take here the opposite point of view: $f$ is large so that the singlet plays the role
of the QCD axion.}

This symmetry breaking pattern can be realized with Sp($n$) TC gauge interactions, 
and the following TC fermion content~\cite{TCSp}
\beq 
\begin{array}{cccccc}\hline
\hbox{Fermions}& {\rm U}(1)_Y&\SU(2)_L&\SU(3)_{\rm c} & {\rm Sp}(n)_{\rm TC} & {\rm U}(1)_{\rm PQ}\cr \hline
D &  \phantom{-}0 & 2 &1 & n & +1\cr
S & +{1 \over 2} & 1 &1 &  n & -1 \cr
 \bar S &-{1 \over 2} & 1 & 1 & n& -1\cr
\end{array}
\label{TCSp}\eeq
Notice that Sp($n$) dynamics induces a condensate that does not break $\SU(2)_L$, despite the presence of a
single lepton doublet.  From the PQ charges it follows that there are no TC-TC-PQ anomalies. 
At this level however $a$ is not a viable axion because QCD-QCD-PQ anomalies are also absent:
indeed when generating fermion masses by adding  four-fermion operators like $(qu)(DS)$,
the SM term and the TC term separately preserve the PQ symmetry. 
One way to remedy this is to generate flavor with dimension 9 operators of the form $(qu)(DS)(S \bar{S})$ so that the SM fields are charged 
under the PQ symmetry. Another way is to consider  non minimal models adding extra TC fermions.

\smallskip

We here consider a different route and assume that fermion masses are generated through linear couplings to states of the strong sector.
This hypothesis, known as Partial Compositeness, is for example realized in extra-dimensional theories, but as drawback
it is not known a microscopic dynamics that realizes it.
Within Partial Compositeness, SM fermions $\psi$ couple linearly to composite fermions $\Psi$, with identical quantum numbers under
the SM group, effectively realizing the following structure:
\beq m \psi \Psi + M \Psi \Psi + g_{\rm TC} \Psi \Psi H.  \eeq
After integrating out the heavy fermions $\Psi$ one obtains Yukawa couplings $(m g_{\rm TC}/M) \psi\psi H$.
Through the mixing terms the SM fields can become charged under the PQ symmetry generating in turn the
required QCD anomaly.

As a simple example we consider the case where SM fermions couple to composite fermions $\Psi_Q, \Psi_U$ in the
{\bf 6} of SO(6).
For the purpose of writing invariants it is useful to promote the SM quarks to incomplete SO(6) multiplets.
For the up sector,
\begin{eqnarray}
\label{embed6}
\psi_{q} &=& \frac{1}{\sqrt{2}} \left(
b_L,  -i b_L ,   t_L , i t_L , 0 ,0\right),\\
\psi_u &=& \left(
0,0,0,0, i \cos\theta\, , \sin\theta \right) u_R
\end{eqnarray}
and similarly for the other fermions, see \cite{singlet,lightpartners} for more details. 
The mixing of left-handed doublets $\psi_q\Psi_Q$ term is automatically invariant under U(1)$_{\rm PQ}$,
while the right-handed one $\psi_u \Psi_U$ is invariant provided that
$\theta=\pi/4$ and $u_R$ has unit charge.\footnote{Alternatively, one can assume that $\Psi_U$ forms a {\bf 10} of SO(6), which contains
a single state with the quantum numbers of $u_R$,  such that U(1)$_{\rm PQ}$ is automatically conserved~\cite{singlet}.}

The one-loop potential for $\Phi$ is \cite{lightpartners},
\begin{equation}
V \simeq C_1\frac{h^2}{f^2} + C_2\frac{h^2}{f^2} \bigg[1-\frac{h^2}{f^2} \bigg]
\label{C1C2}
\end{equation}
where the coefficients $C_i\sim g_{SM}^2  \Lambda^2 f^2/(4\pi)^2$ are generated by gauge and Yukawa interactions.
The  electro-weak vacuum $v\ll f$
is obtained by fine-tuning $C_1+ C_2 \approx v^2  f^2$ to be small
(i.e. by fine-tuning  $g,g'$ and fermionic mixings); 
the correct Higgs mass is obtained from a small $h^4$ term after including the 
SM RGE corrections between $f$ and $v$.
We will come back to this point in the next section.

\medskip

The field $a$ is now a good axion candidate. Indeed the right-handed up and down
quarks are now charged under U(1)$_{\rm PQ}$, with a QCD anomaly
$N=2 N_F$, where $N_F=3$ is the number of generations. The electro-magnetic anomaly is,
\begin{equation}
E=2\left[\left(\frac 4 9+\frac 1 9\right)3+1\right]N_F + E_{\rm TC}
\end{equation}
where $E_{\rm TC}$ is the contribution of the TC fields.
Therefore
\begin{equation}
\frac E N=\frac 8 3 + \frac {E_{\rm TC}} {6}.
\end{equation}
The model of eq.~(\ref{TCSp}) predicts $E_{\rm TC}=2n$ but this is not known in the model with
partial compositeness since we ignore the microscopic theory. 
In general  the electro-magnetic anomaly of the TC sector $E_{\rm TC}$  
grows with the degrees of freedom of the TC sector, $E_{\rm TC}\propto n$, which could be
large.

\section{The Higgs Mass}
\label{higgsmass}
We now discuss the Higgs mass in CH models with large compositeness scale $\Lambda$.
One motivation for such models, independent of axions, is that the Higgs quartic  coupling $\lambda$, normalized such that
$V(h)=m^2 h^2/2 + \lambda  h^4/4$, is predicted to be small at $\Lambda$.  
Present data imply $\lambda(10^{11}\GeV) = -0.003\pm0.02$ (see fig.~\ref{Fig:quartic}a).
The leading contribution is  of order
\beq \lambda(\Lambda) \sim g_{\rm SM}^2
 \frac{g_{\rho}^2 }{(4\pi)^2}  \sim\hbox{few}\,10^{-2}
\label{leadingquartic} 
\eeq
where $g_{\rm SM}$ indicates the various SM couplings $g, g', \lambda_t$ and
$g_\rho\sim \Lambda/f$ is the coupling of the resonances of the strong
sector, $1\lsim g_\rho\lsim 4\pi$. 
In a TC gauge theory $g_\rho$ is given by the large-$n$ relation $g_\rho\sim 4\pi/\sqrt{n}$.  In certain cases (see below) this dominant 
contribution to $\lambda$ cancels out, so that $\lambda$ is even smaller,
generated only by SM interactions: 
\beq \lambda(\Lambda) \sim \frac{g_{\rm SM}^4}{(4\pi)^2}\sim 10^{-3}.\eeq

There are various contributions to the Higgs potential: SM gauge interactions and 
interactions that generate the SM fermion masses. Their sum must be tuned so the electro-weak vacuum $v\ll f$
is generated. We start with the contribution of gauge fields that is essentially universal and can be reliably determined. 

\subsection{The gauge contribution}
To be concrete we will frame our discussion for models
where a single Higgs doublet is present.

\begin{figure}[t!]
\begin{center}
$$\includegraphics[width=0.45\textwidth]{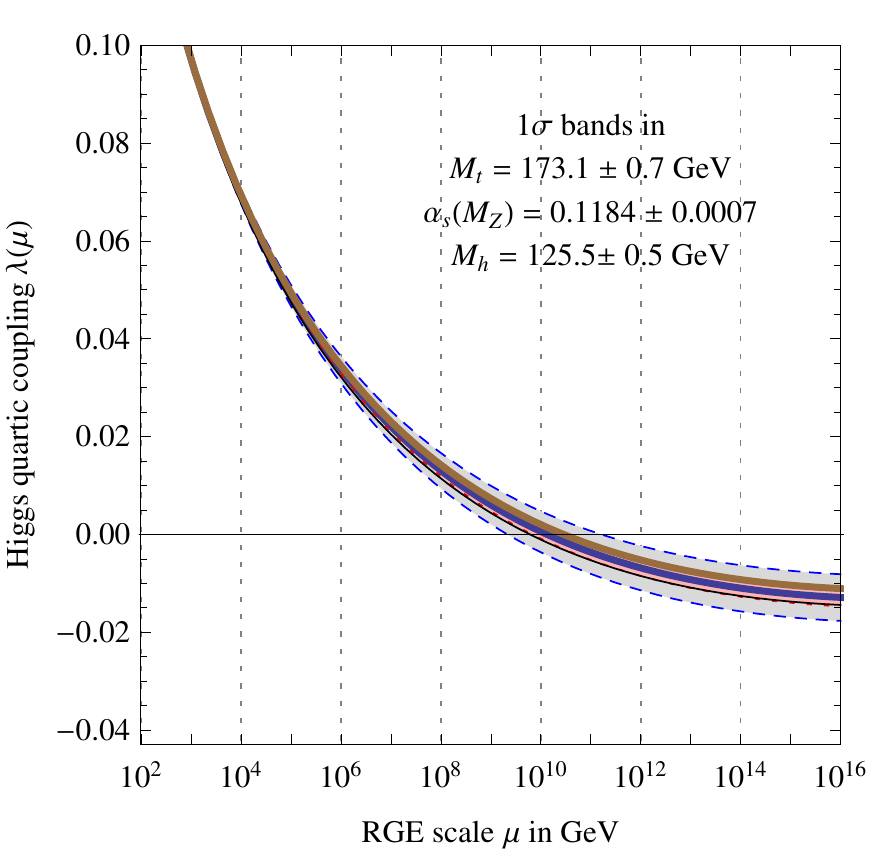}\qquad\includegraphics[width=0.43\textwidth]{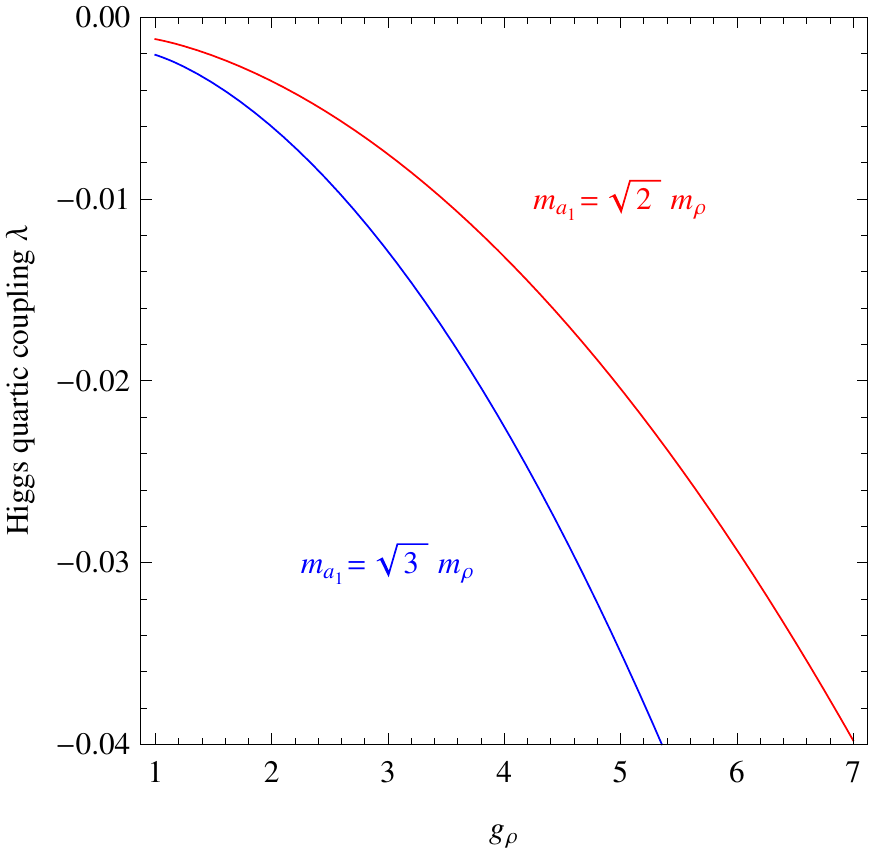}$$
\caption{\em SM RGE evolution of the $\overline{\rm MS}$
Higgs quartic coupling (left) and
gauge contribution to the Higgs quartic coupling $\lambda$ renormalized at the compositeness scale,
as a function of $g_\rho$ (right) for $m_{a_1}=\sqrt{2}m_\rho$ (upper curve) and 
$m_{a_1}=\sqrt{3}m_\rho$ (lower).
}
\label{Fig:quartic}
\end{center}
\end{figure}

The contribution to the effective potential of gauge loops is entirely analogous to the electro-magnetic splitting of
pions in QCD  and is given by the Coleman-Weinberg effective potential,
\begin{equation}
\label{Coleman-gauge}
V(h)_{\rm gauge}= \frac{9}{2}\int \frac{d^4 p}{(2\pi)^4}\ln \left[1 +
F(p^2)\,\sin^2\frac{h}{f} \right]
\end{equation}
where the form factor $F(p^2)$ is expressed in terms of the two point functions of the global currents of the theory,
see~\cite{minimalcomposite} for more details. This can be computed explicitly assuming that the potential is dominated by the lowest resonances,
like in QCD where they are called $\rho$ and $a_1$.

In models with low compositeness scale the momentum integral in eq.~(\ref{Coleman-gauge})
is performed from zero to infinity. 
Since we want to take into account SM RGE effects between the weak scale and the compositeness scale,
we prefer 
to estimate the boundary condition for $\lambda$ at the mass of the lowest resonances ($m_\rho\sim \Lambda$). 
To do this we take a Wilsonian point of view and integrate over momenta
between $m_\rho$ and infinity. 
Expanding the logarithm in eq.~(\ref{Coleman-gauge}) at first order we obtain the leading corrections of eq.~(\ref{leadingquartic}).
More precisely following \cite{4dcomposite} we get
\begin{eqnarray}
\label{explicit-pot-gauge}
V(h)_{\rm gauge}&\approx& \frac{9}{2}\, \left( \int_{m_\rho}^{\infty} \frac{d^4p}{(2\pi)^4} \,F(p^2)\,\right) \sin^2 \frac h f\nonumber \\
&=& \frac{9}{4}  \frac{1}{16\pi^2} \frac{g_0^2}{g_\rho^2} \frac{m_{\rho}^4 \left(m_{a_1}^2-m_{\rho}^2\right)}{m_{a_1}^2-m_{\rho}^2(1+g_0^2/g_\rho^2)}\ln \left[\frac{m_\rho^2+m_{a_1}^2}{m_{\rho}^2(2+g_0^2/g_\rho^2)} \right] \sin^2\frac{h}{f}
\end{eqnarray}
where $m_\rho = g_\rho f/\sqrt{2}$;
$a_1$ are the vectors in $G/H$ that re-establish the global $G$ symmetry and thereby act as an UV cut-off;
$g_0$ and $g_\rho$  are the elementary and composite couplings related to the SM couplings by,
\begin{equation}\label{gaugecoupling}
\frac 1 {g^2}=\frac 1 {g_0^2}+ \frac 1 {g_\rho^2}.
\end{equation}
The formula above includes the effect of mixing with elementary fields that may be relevant in the small $g_\rho$ region. 
The difference with respect to
the integration from 0 to $\infty$ is contained in the logarithm, see \cite{4dcomposite}. 
Numerically for $m_\rho =10^{11}$ GeV the typical reduction is by a factor $\sim 1.7$. To gain some intuition into this formula we 
take $m_{a_1}=\sqrt{2} m_\rho$ similarly to QCD,  finding
\begin{equation}
\lambda(m_\rho)_{\rm gauge}^{\rm leading}\approx - 3g_0^2 \log \frac{3}{2} \frac {g_\rho^2}{(4\pi)^2}   .
\label{leadingquartic2}
\end{equation}
Using the full expression (\ref{explicit-pot-gauge}) one finds the result in Fig.~\ref{Fig:quartic}. A quartic of the order of the 
experimental uncertainty requires $g_\rho \lsim 3$. Note that the gauge contribution to the Higgs quartic is negative at the scale $m_\rho$: 
this does not mean that the full quartic will be negative since other contributions to the potential must exist to tune the 
Higgs mass term to be $m\sim v$ rather than $m\sim f$.
Indeed the gauge contribution to the Higgs mass term is given by
\beq m^2| _{\rm gauge}^{\rm leading} = -\frac{4}{3} f^2 \lambda_{\rm gauge}^{\rm leading}\label{mlambda}\ .\eeq

The TC coupling is expected to be of order $g_{\rho}\sim 4\pi/\sqrt{n}$ for an SU($n)_{\rm TC}$ group,
and thereby $g_\rho$ is small if $n$ is large --- but not smaller than $g$, in view of 
eq.~(\ref{gaugecoupling}). In the example  in section 4 the measurable ratio $E/N$ grows with 
$n$. 

\subsection{The Fermion Contribution and Tuning}

The second unavoidable contribution to the potential is associated to the fermion couplings.
This is model dependent: the SM fermions can couple linearly or bi-linearly to the strong sector 
fields whose quantum numbers under the global symmetries are also not known in general. 

In composite models at the weak scale, the fermionic contribution is often dominant and the cancellation
needed to obtain the electro-weak scale is between different fermionic contributions.
This is no longer necessary in our case because at $\Lambda\approx 10^{11}$ GeV the top Yukawa couplings is numerically similar
to the weak gauge coupling, producing contributions to $m^2$ and $\lambda$ of similar size as the gauge contributions.
Therefore the weak scale can emerge as a huge cancellation in $m^2=m^2_{\rm gauge} + m^2_{\rm fermions}$.
Given the couplings that break the global symmetry of the theory, one could in principle
compute whether such cancellation occurs for the  observed values of the SM gauge and top Yukawa couplings.

If the electro-weak scale emerges through the cancellation of two leading order terms 
with different functional dependence on the Higgs  (as in eq.~(\ref{C1C2}))
the estimate of the  quartic is given by eq.~(\ref{leadingquartic}). 
This is the scenario typically considered in models with low compositeness scale. 
In our context a small quartic then requires a not very large TC coupling $g_\rho$.

\medskip

Another possibility arises: in various models  the leading order 
terms $V(h)_{\rm gauge}$ and $V(h)_{\rm fermions}$ have the same dependence on $h$:  $\sin^2 h/f$. 
The cancellation of the $m^2 h^2/2$ term then implies a cancellation of the whole potential $V(h)$
at leading order in $g_\rho^2$. This is a problem in the context of the usual weak-scale compositeness models,
where sub-leading terms are typically too small to reproduce the observed Higgs mass.
Moreover this increases the tuning of the theory.

Having abandoned naturalness this cancelation becomes a virtue: it just implies that $\lambda\approx g_{\rm SM}^4/(4\pi)^2$ at the compositeness scale.
A non zero $\lambda$ at low energies is then generated by RGE running down to the weak scale, resulting in a Higgs mass of about 125 GeV.
This happens for example in the model of section \ref{Aa}: $V(h)_{\rm fermions}$ is considered in the appendix, 
finding that it is proportional to  $\sin^2 h/f$, like the gauge contribution.
This feature persists at higher orders in $g_\rho$.

\section{Conclusions and Future Directions}\label{concl}
We started to explore models where TC dynamics at some large scale $\Lambda$
spontaneously breaks a global symmetry, 
producing the Higgs and the axion as composite Nambu-Goldstone bosons,
thereby unifying  the Higgs with the axion.
For a scale $\Lambda \approx 10^{11}\GeV$ this scenario has various phenomenological virtues:
\begin{enumerate}
\item the axion solves the QCD $\theta$ problem and provides the observed DM abundance.
Experiments such as ADMX~\cite{ADMX}
are reaching the sensitivity necessary to test axion DM in the Milky Way 
and to measure the $a\gamma\gamma$ coupling, predicted in each model;

\item the observed Higgs mass is generated via SM RGE effects
from a small Higgs quartic coupling at the compositeness scale,
provided that the Higgs mass term is fine-tuned to be of electroweak size.
In many models such fine-tuning implies a cancellation
in the quartic Higgs coupling at the TC scale, such that very small (of order
$\lambda\sim g_{\rm SM}^4/(4\pi)^2$) rather than small (of order $g_{\rm SM}^2 g_{\rm TC}^2/(4\pi)^2$);

\item deviations from the SM are suppressed by powers of $\Lambda$, the dominant effect being Majorana
neutrino masses $m_\nu \sim v^2/\Lambda$.
\end{enumerate}

This scenario predicts no observable deviations from the SM in current precision, flavour and collider experiments.
Like it or not, this is precisely what  experiments found so far.
In this situation, many efforts have been devoted to 
imagining new physics  that  escaped detection, at the price 
of building contrived models without any known UV completion.

Data support the alternative possibility, that no new physics exists at the weak scale, 
motivating scenarios like the one advocated in the present paper.
Simple models that do not make use of elementary scalars could lead to indirectly testable predictions.

\smallskip

In the present context, the next logical step would be to search for a simple theory of the 4-fermion operators
responsible of fermion masses. Given that all effects  beyond the SM are suppressed 
by the large scale $\Lambda\sim 10^{11}\GeV$, one can add extra flavour gauge symmetries broken by the TC dynamics.
The main idea for a complete theory of these operators is Extended TechniColor (ETC).\footnote{However so far we have not found any $\SU(n)$ ETC group with a chiral anomaly-free
representation (such that fermions are massless and the PQ symmetry automatically arises)
that leads to Composite Higgs. We instead find possibly viable models where TC-chiral fermions presumably
break an extension of the SM gauge group, but we do not know how to study their strong dynamics.}

The fine-tuning needed to achieve $v\ll f$ implies a relation between the
SM gauge couplings and the Yukawa couplings.
If a predictive model will be found and its strong dynamics will be computed,
such relation would allow to indirectly test if the 
weak scale really arises in this way.

\paragraph{Acknowledgments.}
We are grateful to Gilberto Colangelo, Rogerio Rosenfeld, Francesco Sannino and
Giovanni Villadoro for useful discussions. This work was supported by the ESF grant MTT8 and by
SF0690030s09 project. 

\appendix

\section{Structure of the Potential}
\label{Appendix}

We here  discuss the structure of the potential in the models of section \ref{Aa} and \ref{Ba}, following
the discussion in Ref. \cite{luty}.

\subsection{Potential in the $\SU(6)_L\otimes \SU(6)_R/\SU(6)_{L+R}$ model}
The NGBs are described by an $\SU(6)$ matrix $U$. 
The gauging of SM group generates a potential
for charged states analogous to the electro-magnetic splitting of the pions. It can be estimated as,
\begin{equation}\label{A1}
V_{\rm gauge}\sim \frac 3 {16\pi^2}\ \Lambda^2 f^2 \sum_a g_a^2 {\rm Tr}[ U T^a U^\dagger  T^{a}]
\end{equation}
where $T^a$ are the SM  gauge generators. This generates a positive contribution to the squared mass of all the charged states. 
By assumption only the Higgs will be however light due to cancellation with different contributions to the potential.

Let us now turn to the fermions. 
Adding 4-Fermi operators one obtains a top Yukawa term
\begin{equation}
 C_t\, \sum_{\alpha=1}^2 (q_L^\alpha t_R^c)^\dagger \,{\rm Tr}\, [\Pi^\alpha_{t}\cdot U]+\hbox{h.c.}
\label{yukKSVZ}
\end{equation}
where $C_t\sim \Lambda f^2/\Lambda_t^2\sim f\lambda_t$. The matrices $\Pi_{t}$ are  determined by the combination that couples to the SM fermions in 
eq.~(\ref{yuk}), given by
\begin{equation}
\Pi_t^1=\frac 1{\sqrt{|a|^2+|b|^2}}
\left(\begin{array}{cccccc}
0 & 0 & 0 & 0 & 0 & 0\\
0 & 0 & 0 & 0 & 0 & 0\\
0 & 0 & 0 & 0 & 0 & 0\\
0 & 0 & 0 & 0 & 0 & 0\\
0 & 0 & 0 & 0 & 0 & a\\
0 & 0 & 0 & 0 & b & 0\\
\end{array}\right)\,,~~~~~~~
\Pi_t^2=-\frac 1{\sqrt{|a|^2+|b|^2}}\left(\begin{array}{cccccc}
0 & 0 & 0 & 0 & 0 & 0\\
0 & 0 & 0 & 0 & 0 & 0\\
0 & 0 & 0 & 0 & 0 & 0\\
0 & 0 & 0 & 0 & 0 & a\\
0 & 0 & 0 & 0 & 0 & 0\\
0 & 0 & 0 & b & 0 & 0\\
\end{array}\right)
\end{equation}
where the numbers $a,b$ depend on the specific 4-fermion operator considered.

At 1-loop the leading contribution to the potential can be estimated as
\begin{equation}
V_{\rm fermions}\sim \frac{N_c\,\lambda_t^2}{16\pi^2}\Lambda^2 f^2 \  \sum_{\alpha=1}^2 \left| {\rm Tr} [\Pi^\alpha_t\cdot U] \right|^2.
\label{VfermAa}
\end{equation}
This term has a $\sin^2 h/f$ dependence on the Higgs field $h$, like the term in eq.~(\ref{A1}) generated from  gauge interactions.

\subsection{Potential in the $\SU(6)/\SO(6)$ model}\label{A2app}

To describe the low energy interactions with this pattern it is useful to observe that the breaking 
can  realized through the VEV of the symmetric representation of $\SU(6)$. 
The NGBs can be parametrized by
\begin{equation}
\Phi = U\cdot  \Phi_0 \cdot U^T\,
\label{DFSZvev}
\end{equation}
where $\Phi_0$ is any vacuum expectation value that breaks $\SU(6)\to \SO(6)$
(for example $\Phi_0={\rm diag}\,(1,1,1,1,e^{i\alpha},e^{-i\alpha})$)
and $U= e^{i  {\pi^{\hat{a}} T^{\hat{a}}}/f}$ where the sum extends to the broken generators.   
The low energy lagrangian is then simply obtained by writing formally $\SU(6)$ invariant terms
containing $\Phi$.

The gauge contribution to the potential is now
\begin{equation}
V_{\rm gauge}\sim \frac 3 {16\pi^2}\ \Lambda^2 f^2 \sum_a g_a^2 {\rm Tr}[ \Phi T^a \Phi^\dagger  T^{a*}]
\end{equation}
with identical properties to the one discussed above.
The 4-Fermi operators (\ref{dim6DFSZ}) generate
\begin{equation}
{\cal L}=  C_t\, \sum_{\alpha=1}^2 (q_L^\alpha t_R^c)^\dagger \hbox{Tr}\, [\Pi^\alpha_t\cdot \Phi]+
C_b\, \sum_{\alpha=1}^2 (q_L^\alpha b_R^c)^\dagger  \hbox{Tr} \,[\Pi^\alpha_b\cdot \Phi]+{\rm h.c.}
\label{yukDFSZ}
\end{equation}
where the matrices  associated to the top Yukawa are
\begin{equation}
\Pi_t^1= \frac 1 2
\left(\begin{array}{cccccc}
0 & 0 & 0 & 0 & 0 & 0\\
0 & 0 & 0 & 0 & 0 & 0\\
0 & 0 & 0 & 0 & 1 & i\\
0 & 0 & 0 & 0 & i & -1\\
0 & 0 & 0 & 0 & 0 & 0\\
0 & 0 & 0 & 0 & 0 & 0\\
\end{array}\right)\,,  \qquad
\Pi_t^2=\frac 1 2\left(\begin{array}{cccccc}
0 & 0 & 0 & 0 & 1 & i\\
0 & 0 & 0 & 0 & -i & 1\\
0 & 0 & 0 & 0 & 0 & 0\\
0 & 0 & 0 & 0 & 0 & 0\\
0 & 0 & 0 & 0 & 0 & 0\\
0 & 0 & 0 & 0 & 0 & 0\\
\end{array}\right)
\end{equation}
and similarly for the bottom sector. 

The potential generated by quark loops at leading order is given by
\begin{equation}
V_{\rm fermions}\sim \frac{N_c\,\lambda_t^2}{16\pi^2}\Lambda^2 f^2 \sum_{\alpha} \left| \hbox{Tr}\,  [\Pi^\alpha_t\cdot \Phi] \right|^2+ \frac{N_c\,\lambda_b^2}{16\pi^2}\Lambda^2 f^2 \sum_{\alpha} \left| \hbox{Tr}\,  [\Pi^\alpha_b\cdot \Phi] \right|^2.
\label{VfermBa}
\end{equation}
The bottom contribution could be relevant depending on $\tan \beta =v_1/v_2$. Expanding the top contribution 
around the field origin 
one finds
\begin{equation}
V_{\rm fermions}\sim \frac{N_c\,\lambda_t^2}{16\pi^2}\Lambda^2 \left( |H_u|^2+\frac{2\sqrt{2}}f \, {\rm Im}\,[ \sigma H_u H_d]+\dots\right).
\end{equation}

It is easy to see that the potential for $\sigma$ is controlled by the Higgs vacuum expectation value and is therefore suppressed in the electro-weak vacuum. 
This however is an accident of the one-loop approximation and at two-loop an unsuppressed potential for $\sigma$ will be generated
that must be taken into account.

The singlet $\eta$ is an exact NGB as long as SM couplings are considered.  This is phenomenologically allowed
in view of $\Lambda \approx 10^{11}\GeV$.
Anyhow, it can be made massive
by adding to the Lagrangian
mass terms for the TC-fermions 
that break explicitly the symmetry associated to $\eta$
and respect the
SM gauge symmetry and the PQ symmetry:
\begin{equation}
m_L L {\bar L}+ m_N\, N \bar{N}+\hbox{h.c.}
\end{equation}
Similarly to quark masses in QCD, this mass matrix $m$ generates
\begin{equation}
V_{\rm mass}=\Lambda\, f^2\, {\rm Tr}[ m \cdot \Phi] + \hbox{h.c.}\simeq
 \Lambda m_L (12 |H_u|^2+12 |H_d|^2+ \eta^2)+ \Lambda m_N (6|H_u|^2+ 6|H_d|^2+ 6|\sigma|^2+\eta^2)+\dots
\end{equation}
To realize the DFSZ axion model, $\sigma$ must spontaneously break the PQ symmetry
getting a vacuum expectation value, which (in absence of tuning) 
is of order  $f$. This corresponds to $\alpha\sim \pi/2$ in eq. (\ref{DFSZvev}). Around this vacuum the Higgs potential must be
tuned to reproduce the electro-weak  vacuum $v\ll f$.

\end{document}